# Openness in AI and downstream governance: A global value chain approach

Working paper, 12 September 2025[1]

Christopher Foster, University of Manchester, UK

christopher.foster-2@manchester.ac.uk

## Abstract

*The rise of AI has been rapid, becoming a leading sector for investment and promising disruptive impacts across the economy. Within the critical analysis of the economic impacts, AI has been aligned to the critical literature on data power and platform capitalism – further concentrating power and value capture amongst a small number of "big tech" leaders.*

*The equally rapid rise of openness in AI (here taken to be claims made by AI firms about openness, "open source" and free provision) signals an interesting development. It highlights an emerging ecosystem of open AI models, datasets and toolchains, involving massive capital investment. It poses questions as to whether open resources can support technological transfer and the ability for catch-up, even in the face of AI industry power.*

*This work seeks to add conceptual clarity to these debates by conceptualising openness in AI as a unique type of interfirm relation and therefore amenable to value chain analysis. This approach then allows consideration of the capitalist dynamics of "outsourcing" of foundational firms in value chains, and consequently the types of governance and control that might emerge downstream as AI is adopted. This work, therefore, extends previous mapping of AI value chains to build a framework which links foundational AI with downstream value chains.*

*Overall, this work extends our understanding of AI as a productive sector. While the work remains critical of the power of leading AI firms, openness in AI may lead to potential spillovers stemming from the intense competition for global technological leadership in AI.*

---

[1] As earlier version of this paper was presented in the AI value chain workshop in Kings College, London in May 2025 and the GVC stream at SASE 2025, Montreal, Canada, July 2025 with thanks to feedback from organisers and attendees.

# 1. Introduction

The rise of AI has been rapid, becoming a leading sector for investment and promising disruptive impacts across many sectors of the economy. Within the critical analysis of the economic impacts, AI has predominantly been aligned with the literature on data power and platform capitalism, with risks around value capture amongst a very small number of "big tech" leaders who build major AI models.

The focus of this paper is on the equally rapid expansion of openness in AI, here taken to encompass a set of claims made by firms about AI openness, "open source" and free provision[2]. The extent of this shift has been significant. From the early days of AI, many open resources have been present, including open libraries (e.g. PyTorch, TensorFlow), key data sources (e.g. Common Crawl) and early AI models (e.g. BERT). However, debate on openness has particularly come to the fore as general-purpose AI models have been opened. Notable cases include Meta's – with a reported $60-$65bn capital expenditure on AI, major outputs such as the LLaMA model released under a relatively permissive open licence. This implies that, with some limits, the LLaMA model can be downloaded and commercially reused without a fee, with flexibility around hosting and model fine-tuning (adapting the model). Similarly, the emergence of Chinese models, such as Deepseek-R1 and Qwen3 (Alibaba), with models licenced under near open-source conditions, signifies an acceleration of these trends in the Chinese AI industry.

This rise in openness signals an interesting development in how AI models are used and applied in broader industries. Advocates suggest that such open resources can support technological transfer. In the longer term, openness may facilitate the potential for firms to "catch up" through technology learning, even in the face of AI power (Lee 2018). In contrast, more critical views have questioned the motivation for large firms to be open. Critiques highlight the continuing gaps in terms of AI transparency and ethics, as well as the continuing appropriation of software and data commons (Widder et al. 2024). From this perspective, openness is recast as "open washing", which cements the power of leading AI firms.

This work seeks to add conceptual clarity by integrating these two debates – on openness and power. It asks the following question: *with the emergence of openness discourses in AI, how*

---

[2] We use the generic terms "openness" and "open" in this paper to describe firm actions. These terms are used rather than the more common "open source AI" to emphasise that the formal definition of open source in AI is still highly contested, and some firms would not be considered to meet such definitions. See elaboration in section 2.

*should we conceptualise the major impacts of openness on power and agency?* To answer the question, the paper extends previous work on AI that has sought to map AI production and use, often highly fragmented, within a framework of "AI value chains" or "AI supply chains". This paper develops these mappings to think critically about the trade-offs between leading firm power and the opportunities for AI users in the era of "open" AI. It does this through a critical engagement with theories from global value chains (GVC) and emerging empirical data around openness in LLM models.

Through this analysis, the paper develops a conceptual model of AI value chains. We develop the notions of *strategic market openness* to explain the motivation of foundational AI firms for openness and outline the resultant *heterogeneous governance* that posits different patterns of relations with firms downstream in AI chains. Overall, while the work remains critical of the power of leading AI firms, openness in AI offers potential spillovers stemming from the intense competition for global technological leadership in AI. While some still see debates on openness AI as a niche or temporary aspect of AI development, the notion of *strategic market openness* also suggests that there are fundamental dynamics around openness that are liable to remain key to AI and even expand in the future.

The significance of this paper is that it considers more systematically how paradigms of openness, often set aside from core analysis of the economics of AI, link into broader critical debates on AI. This includes its impacts on capabilities, firm governance and economic outcomes. Previous work on AI power has tended to be concentrated at either a macro-level to focus on power or take a purely empirical approach to understand specific AI firms. The adaptation of GVC frameworks for AI allows the development of a *meso-level framework* for the analysis of foundation AI firms linked to openness. This approach also provides policy-relevant knowledge. As GVC frameworks have previously shown, firms downstream in value chains look to economically upgrade and specialise, centred on the building of substantial relationships with lead firms, facilitating learning and technology transfers. The conceptualised model here, therefore, provides reflection on future capabilities and upgrading pathways in the face of power in the AI value chain.

The remainder of the paper is set out as follows. Section 2 provides a more detailed overview of major debates around openness and AI. Previous literature on "AI value chains" and "AI supply chains" is also reviewed, where critical questions around openness have primarily been positioned in terms of AI transparency, ethics and security. Based on the above, in section 3, the global value chain (GVC) framework is analysed in the light of AI. This supports the development of an analytic frame for this analysis, particularly centring on three major concepts in GVC – software/services in value chains, the dynamics of capitalism and governance. In section 4, three major concepts are applied to the case of openness in LLM models. Notably, we examine the motivations for foundational AI firms to externalise key components of AI production through openness and how externalised AI components mediate relationships with downstream firm users of AI. Developing this, in section 5, a model of heterogeneous AI governance and openness is developed.

## 2. A relational approach to openness and AI

### 2.1. Open source and AI

Early releases of major AI models have attracted major attention, with focus centred around the leading firms with multi-billion dollar AI investments, such as the likes of OpenAI, Anthropic and Meta. Building AI models has been seen as a capital-intensive activity with the need for significant outlay on R&D, technical skills, data resources, data transformation, and AI compute to build models which operate at the leading-edge (Srnicek 2020). Accordingly, much of the AI scholarship has focused on these handful of firms where previous critical discussion of power in digital capitalism and the emergence of big tech has been extended to the concentrated AI industry[3] (e.g. van der Vlist et al. 2024, Widder et al. 2024). These dynamics are reflected in the popularisation of the terminology of "foundational models" which describes the all-encompassing and concentrated trajectory of AI, *"[a] foundation model is any model that is trained on broad data (generally using self-supervision at scale) that can be adapted (e.g., fine-tuned) to a wide range of downstream tasks..."* (Bommasani et al. 2022 p.3). Foundational models and their firms have been seen as important to focus on because they "...*have the potential to accentuate harms, and their characteristics are in general poorly understood. Given their impending widespread deployment, they have become a topic of intense scrutiny"* (Bommasani et al. 2022 p.3).

Before discussing openness and AI, it is useful to position openness within the longer-running debates on the digital economy. From the early days of the internet, the production of "open source" software has been a significant aspect of the field, with Linux (operating systems), Apache (web server) or Python (programming language) being frequent examples cited. The emergence of open source inspired economists and business scholars towards predicting these outputs as indicative of potential new forms of collaboration, exchange and inclusive economy that might have broad implications for capitalist trajectories (Benkler 2006, Lessig 2004). However, notions of commons-based peer production and peer-to-peer production have been slow to emerge. Rather, there has been integration between open source and private firms - to the extent that most highly capitalised and profitable digital services (including major cloud and platform firms) are often powered by open source infrastructure. Major ideas within the business literature have, therefore, evolved to make a closer consideration of open source within dominant capitalist paradigms. Notable work includes analysis from innovation studies, examining the links between open source communities and firm-level innovation that lead to new sources and forms of "user innovation" (Hippel & Krogh 2003, Von Hippel 2001). Parallel notions of openness have emerged in the business strategy discussion of "open innovation", which centres open paradigms as new models for inter-firm "co-creation" (Chesbrough 2006). Some literature has also taken a more critical standpoint. Rather than economic transformation, private firm appropriation of open source is argued to simply lead to private profit on the back of free "labour", intellectual property and privatisation of the digital commons - simply another turn in late capitalism (Stalder 2018).

---
[3] Not least because all the leading platform firms are key owners or investors within AI firms.

It is within these debates that what is often called "open source AI" has emerged. Given AI's long history as an academic research area, it is unsurprising that many well-recognised AI tools and resources are available publicly as the outputs of publicly supported research (Wooldridge 2021). Alongside this, a growing range of smaller-scale open source AI projects are side-projects of professional AI developers and enthusiasts in this area. However, the step-change has been the emergence of large-scale AI models, with claims around openness, including from the likes of Meta and Deepseek. Here, huge capital investments made into developing foundational AI models are being shared openly.

These latter trends have prompted renewed discussion on openness. Many of these debates have concentrated on openness linked to AI ethics, safety, transparency, replicability and privacy. Debate has pondered on questions of the risks of the open vs closed paradigms for AI, in terms of security and ethical implications (Gent 2024). More formally, open source definitions have been adapted for a definition of open-source AI that would maximise availability and transparency,

> "An *Open Source AI is an AI system made available under terms and in a way that grant the freedoms to:*
> - ***Use*** *the system for any purpose and without having to ask for permission.*
> - ***Study*** *how the system works and inspect its components.*
> - ***Modify*** *the system for any purpose, including to change its output.*

- **Share** *the system for others to use with or without modifications, for any purpose."*
  *(OSI n.d. emphasis from original)*

Openness across all dimensions of this definition would be important if it is to maximise societal gains (e.g. allowing clear analysis of AI decision making, fair use of data and building replicable and reliable models) (Larsson & Heintz 2020). In contrast to this definition, the current reality is that many AI firms only make claims along some of the above dimensions, and under specific conditions. These conclusions have led to arguments that many claims of "open source" should be seen as "open washing", allowing AI to appear more ethical and societally beneficial, without adhering to what would be required for this in reality (Widder et al. 2024). Open source debates have, therefore, provided a foundation for a normative evaluation of "responsible AI systems" in practice.

While these discussions frequently allude (and somewhat overlap) with the discussion of the economics of AI, a broader discussion of the economic trajectories linked to openness has only been touched upon in a few discussions (e.g. IHCAI 2023). Notwithstanding the broader ethical and risk issues, from an economic perspective, the link between the extent of open AI and outcomes may depart somewhat from the debate on open source AI. This is because even partially openly available resources are significant for others in terms of mediating relationships of technology use, including potential technology transfer, cost reduction and skill building. As Kai Fu Lee suggested in 2018, predicting the later rise of Chinese AI, broader types of openness can be beneficial because, *"...the center of gravity quickly shifts from a handful of elite researchers...Facilitating this knowledge transfer are two defining traits of the AI research community: openness and speed''* (Lee 2018 p.86). The implication is that even with constraints, openness may allow actors and firms outside leading firms the ability to rapidly learn and potentially compete, even within the massively capitalised AI startup sector.

In sum, the emergence of openness in AI should not be seen as separate from the broader debates in the digital economy. As the history of debates on open source outlined in this section shows, there are different positions on the way that AI openness might link to socio-economic outcomes, including the expansionary potentials of technological learning and innovation and the societal risks of "openwashing" and control in foundational models.

## 2.2. Chain approaches to AI

There have been a number of attempts to draw on relational approaches to analyse the AI expansion. These respond to the call to consider the wider implications of AI industries, not least to provide perspectives on the debates outlined in the previous section. As Crawford(2021

p.11) puts it *"We need a theory of AI that accounts for the states and corporations that drive and dominate it [AI], the extractive mining that leaves an imprint on the planet, the mass capture of data, and the profoundly unequal and increasingly exploitative labor practices that sustain it"*. Aligned closely to Crawford's call, some relational analysis resemble the previous approach to "commodity chain" analysis, building analysis of relations and division of labour for the case AI (Bair 2005). Taking inspiration from sociological analysis that has conceptualised the emergence of planetary computation (Bratton 2016), this critical work seeks to outline the ways that AI envelopes, appropriates or shapes production activities across a broader array of production - both upstream and downstream, both tangible and intangible, both directly and indirectly (Ferrari 2023, Hung 2024, van der Vlist et al. 2024). Such approaches have taken different scales. At an application level, examining services centred on AI infrastructure, van de Vlist et al.(2024) outline the dynamics of expansion of big tech firms, which power the infrastructure used in AI, with significant impacts in terms of uneven power across business ecosystems. At a more societal level, drawing on notions of supply chain capitalism, other literature provides insights on the material relationships and division of labour that are reproduced and enhanced as AI emerges, such as through highlighting the centrality of critical mineral production and the sites of data labour (Crawford 2021, Valdivia 2024).

Another (complementary) motivation for taking a relational approach to AI would be Mueller's provocative arguments regarding the imprecise definitions of AI. In reality, to understand AI is to explore how different components of distributed computing – devices, networks, data and software – come together within contemporary systems that we now call "AI" or "machine learning" (Mueller 2025). From this view, an important step for economic analysis is to detail the value distribution within these processes. Developing earlier work on the data value chain (Foster & Azmeh 2023, Gurumurthy et al. 2020), AI "value chain" analysis has begun to map the major processes at different scales including: the chains for creating and using machine learning models, the broader chains of computing and infrastructure resources that are part of AI, and the ways that AI interact with other sectoral value chains (Gambacorta & Shreet 2025, Heeks & Spiesberger 2024). Using well-established value chain analysis allows a more systematic analysis of the different actors and relations, including better revealing the labour involved across the AI chain (Anwar 2025). Given the broader ethical debates on AI, mapping value stages has also been used as a methodology to examine ethics and societal aspects of AI. Specifically, analysis of AI ethics (including emergent AI policy) needs to account for the distributed chain of actors involved in production and therefore reflect on issues of responsibility, cybersecurity and regulatory gaps as AI traverses multiple stages of production (Attard-Frost & Widder 2025, Brown 2023, Herpig 2020). Broadly speaking, and with some exceptions (Küspert et al. 2023), the focus of "AI value chains" has tended to be on the upstream activities in AI. That is, the focus is on how different data and infrastructure are harnessed in the production of AI models, as opposed to the details of AI use as it is adopted within industries[4].

The above literature reflects how chain approaches (more aligned to commodity or value chains) have been adopted with business, regulatory and social science analysis. However, chain approaches have also been embraced by more technical analyses of AI fragmentation. These approaches tend to be positioned as an AI supply chain analysis (i.e. focuses on the movement of AI technical resources, as opposed to analysis of value distribution or division of labour)[5]. Here, a chain framework is useful to unpack the more complex configuration of AI. For example, foundational AI models in AI may be used by a range of partners to run services (e.g. translation services). These services may then be offered through broader applications or implementations (e.g. a cloud service which includes translation). Supply chain analysis can then track the "bundling" of technical aspects within the AI ecosystem (Cobbe et al. 2023, Hopkins et al. 2023). The extent of such bundling of AI resources is revealed in studies such as the "Ecosystem Graphs" project, a dataset which provides an interesting attempt to map technical supply chains linked to major foundation models from 2023 (Bommasani, Soylu, et al. 2023). This work is insightful for the discussion of open source because it also maps the complex ways that open and proprietary toolsets, libraries and data are combined in foundational models. In turn, these are then supplied (in a more or less open way) to those creating more specific AI applications.

In sum, the emergence of commodity, supply and value chains approaches confirms chain approaches as useful analytical frameworks to explore AI. It highlights some important aspects of understanding value creation in AI. The technical supply chains work, in particular, is valuable in that it highlights the close attention needed to the software, services and integrative logic which mediates all relationships in AI, *"...the value of individual processes in the market becomes secondary to a technical or functional understanding of how these processes work*

---

[4] Although there is a growing complementary literature on algorithmic governance and algorithmic management in work (e.g. Dupuis 2024, Krzywdzinski 2017, Mills 2024, Woodcock 2020, Zhang et al. 2025) this work has been less clearly orientated towards chain structures and also tends to incorporate a range of definitions of algorithmic governances that may include not only AI but also a broader set of data and computationally driven work activities.

[5] Some of this work may use the terminology of the "AI value chain", however this tends to take limited conceptual ideas from value chains and as such they tend to align more closely to technical analysis of supply (for a fuller discussion see Hopkins et al. 2024)

*together"* (Hopkins et al. 2024). However, these are still significant gaps. Most work tends to underplay foundation conceptual ideas from Global Value Chain and Global Production Networks, such as governance, and consequently, the broader economic dynamics have been underplayed in current AI chain analysis. Relatedly, work has concentrated on the upstream configurations of how models are being put together, which, although vital at this early stage of AI, has led to an underplaying of the dynamics of foundational models in their use across industries. Developing a GVC framework for AI and openness the paper, can strengthen how conceptual approaches to AI value chains are undertaken.

## 3. Examining GVC in the context of AI value chains

The review of AI and chain approaches cements the importance of chain frameworks to explore AI and openness. Yet, this literature is still conceptually underdeveloped. This section then makes a more detailed analysis of global value chains in the context of openness and AI.

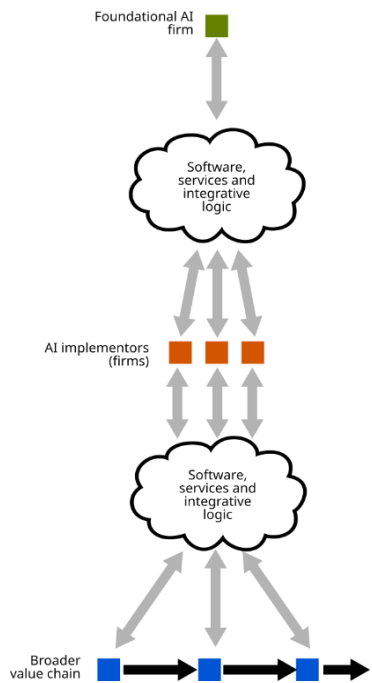

*Figure 1: Simplified model of a downstream value chain of AI. Authors conception*

A starting point here would be a simplified downstream model of the value chain of AI (Figure 1) influenced by the literature from the previous section. At the core of coordination and power are "foundational AI firms", the concentrated producers of foundational models (and/or other major

AI components[6]) with some type of governance, coordination or power over broader chains that use AI. Foundational AI leaders are likely to be in relationships with "AI implementors", likely to be firms that use and/or adapt foundational AI for specific applications/data. AI implementors might include, for example, service providers in an industry who are more likely to interact with the broader downstream value chain.

In line with discussions of cross-cutting aspects of GVC such as finance, information and infrastructure (Dicken 2011), the AI chain is positioned as a vertical chain in contrast to the horizontal sectoral value chain. This indicates that foundational AI activities may be created outside sectoral structure and can cut across multiple industries. A key final point is to emphasise that major relationships between actors in AI chains will be predominantly mediated via software, services or other integrative logic as shown in the clouds in Figure 1.

Although this model represents a stylised model[7], it serves as a starting point to discuss GVC conceptualisations in three areas. Firstly, in terms of the *AI services, software and assets*, it would be important to understand these in terms of the mediating relationships in the value chain. Secondly, in terms of *foundational AI firms*, if these firms have some types of power or governance over value chains, then it is important to understand more deeply their motivations for AI openness. Finally, in terms of *value chain governance*, to reflect on power, coordination and its implications for learning, capabilities and upgrading. In the rest of the section, these three areas are covered in turn.

## 3.1. Value chains of AI assets

Examining concepts of service-dominant production is appropriate for a study of AI value chains (Attard-Frost & Widder 2025). GVC has been dominated by analysis of commodities and global supply chains of goods in the past, and software and services require further consideration. The grounding of GVC in an analysis of value distribution in production relationships does not preclude services even if they represent different dynamics in terms of their production, skills and labour (Gereffi & Fernandez-Stark 2010, Low 2013, Zysman et al.

---

[6] This is a departure of the typically discussion of "foundational AI". In the AI value chain we acknowledge that infrastructure and/or AI-intensive service suppliers (if different to firms producing AI models) may also be highly concentrated and have governance power in GVC. In this section, we will use the term "foundational AI" firms for brevity to indicate this broader definition.

[7] For example, there may be cases where firm relationships are mediated via contract or direct actions in the AI value chain as well as software logic, see next sections. In some industries, it may also be the case that firms in sectoral value chain directly use foundational AI rather than going through intermediaries (e.g. multinational lead firms or lead firms with large R&D departments).

2011). Notably, with software-driven services being highly dynamic, the makeup of services can change more rapidly (Baines et al. 2009).

Within recent conceptualisation of value chains, the combination of service and digitalisation has tended have be associated with the deepening of *modularisation,* which facilitates sharing of complex requirements across networks (Foster et al. 2018, Sturgeon 2021). This highlights the ability for an increasing complexity of goods and services to scale (and therefore chain fragmentation) defined by modularity. In terms of exchange, software and services interlink increasing numbers of actors, often making chains more complex and multi-sectoral (Humphrey 2018, Thun et al. 2022). Therefore, it is the complex standards as set by modularised systems that become important in shaping interfirm relations, whose trade then is increasingly automated by digital systems (Baglioni et al. 2021). Beyond the core ideas of modularity and resultant dynamism, Yoo et al(2010) highlight that the malleability of modular software and services (e.g. algorithms, code) will often lead to flexibility in how production networks are configured. As he highlights for the example of Google maps, this leads to what he calls a more "generative" structure of production network when it is digital service intensive.

> *"…Google Maps can be used as a standalone product, it can simultaneously be used in a variety of different ways, bundled with a host of heterogeneous devices such as desktop computers, mobile phones, televisions, cars, navigation systems, or digital cameras…a product is inductively enacted by orchestrating an ensemble of components" (Yoo et al. 2010 p.728).*

This notion of generativity has important implications because it implies that AI solutions (e.g. AI models, relevant datasets, etc) will frequently be bundled and repurposed across different industries and segments - in ways that might not have been considered by designers, implying potential in adaptation or reuse.

When bringing in openness to this discussion, it poses questions about how relationships are mediated via open software, services and integrative logic. Core to previous global value chain frameworks is that relationships were typically assumed to be transactional, occurring between interconnected firms (i.e. one actor supplying a good or service while the other provides a payment). This assumption needs to be loosened when considering open software networks to include both:

- Traditional GVC transactions – i.e. contract, service agreements or payments.
- Free software or services which nevertheless mediated between the supplier and user.
- Other aspects of contracts that mediate relationships - open-source users may agree to licensing terms, conditions or open-source contracts.

- Firms may indirectly pay for services through their data or actions becoming aggregated or monetised by the provider of services.
- Software, services or integrative logic may have "institutional" power, in the sense that it embeds rules and norms which will shape how implementation can occur (e.g algorithms).
- Informal norms around how to use technology related to industry or broader norms may be relevant (e.g. AI ethical norms)

This openness aspect should broaden our perspectives of lead firm actions, the possibility of appropriation, and the notion of governance within AI value chains.

## 3.2. Openness and capitalist dynamics for foundational leaders

A core aspect of GVC analysis is to outline major trends around fragmentation in global production - looking to examine the emergence of interconnected chains of production and subsequent division of labour, value and relations (Gereffi et al. 2005, Kaplinsky & Morris 2001). A major driver for fragmentation and offshoring of production was originally seen to be centred on multinationals focusing on their "core competencies", maximising their capture of value and outsourcing those stages less central (or of low value) to their business goals (Prahalad & Hamel 1990). With improved global infrastructure and trade, globalisation offered new opportunities for "lead firms" to manage fragmented production in globalised networks (Gereffi 2001).

"Core competencies" remain a major notion (even an assumption) around why fragmentation of production occurs. However, perspectives have been refined over time, most notably within economic geography debates of Global Production Networks (GPN), which seek to bring a stronger (multi-scalar) spatial perspective on such fragmenting production (Henderson et al. 2002). In their later elaboration of GPN, Yeung and Coe (2015) focus on "capitalist dynamics" to better understand lead firm strategy that shapes fragmentation. They outline three fundamental drivers: *cost-capability ratio*, the need to optimise costs but within the contexts of available actor capabilities; *market development*, in terms of the conditions for firms to expand and/or dominate particular markets; and *financial discipline* aligned to the growing financialization of lead firms as a driver of strategies of fragmentation.

These dynamics of capitalism are closely linked to how "value" is distributed in production. That is, lead firms will seek to organise networks of production according to how to maximise value within particular moments within the capitalist lifecycle (Havice & Pickles 2019). As key drivers of wealth creation and profit evolve in the digital economy, it would therefore be important to more fully integrate a discussion of value as a way to think about contemporary trends of fragmentation (Foster 2024). For example, major value trajectories in the digital economy have centred on firms heavily investing to support future financial valuation, keeping control of key data resources and platforms, proprietary algorithms or network graphs, and these may then lead to distinct dynamics of capitalism (Foster 2024).

These changing dynamics can aid consideration of openness and AI. Here, fragmentation equates to the way AI models and major toolkits, amongst others, are shared outside lead firms (often for free). Here classic GVC notion of "core competencies", which mainly focused on firms outsourcing "low value" or labour-intensive activities, appears less fitting. One might conceivably extend this view to consider that "competencies" may also but firm capital (such as compute and data) which AI firms tend to hoard. Similarly, GPN "dynamics of capitalism" potentially provides some explanatory power to consider AI openness, such as the link of financialization and the need to rapidly build AI markets in a possible "winner takes all" scenario. Nevertheless, further work is needed to conceptualise trends of AI openness, which are major drivers of fragmentation.

This discussion is important because, as decades of analysis of outsourcing in GVC have identified, outsourcing is not just about control; it can lead to opportunities for firms. Although *"the functions outsourced by lead firms were 'low profit' at the time this happened. They became more profitable as a result of subsequent technological changes that created new opportunities for scale economies"* (Ponte & Gibbon 2005). Although these arguments also need to be refined for the changed perspectives of AI, these ideas may have further utility in the case where "high value" aspects of AI production are being opened (and hence outsourced).

## 3.3. Conceptualising openness and governance

The standard approach to think about coordination, power and relations in GVC has been Gereffi et al's (2005) conceptualisation of five major types of *governance* in production: ranging from market-based relations, to modular relations, to thicker relational interactions, to supplier captive relations, to vertical integration. This approach to governance has been useful to outline a heterogeneous set of governance relationships where the *"....goal is to bring order to the variety of network forms we see in the world"* (Gereffi et al. 2005 p.79).

The justification underlying these forms of governance was originally conceptualised through reference to transaction cost economics, arguing that governance will predominantly be an outcome of the nature of sectoral properties in terms of the major aspects of transaction: complexity of transactions, codifiability of information and capability of suppliers. With governance at the forefront of GVC analysis for many years, a number of modifications to such models of governance have been suggested to incorporate the diversity of forms of coordination and control linked to lead firms (Horner & Nadvi 2018, Ponte & Sturgeon 2014). This includes proposals that move away from conceiving governance within inter-firm relations to consider how institutions, norms, and technology are important mediating forces in governance across the production chains (e.g. Dallas et al. 2019, Foster & Graham 2017, Ponte & Gibbon 2005, Ponte & Sturgeon 2014). AI and openness analysis should seek to align with these latter developments because (as outlined in 3.1), openness redefines production relationships away from transactions. There is an exchange, but the guarantees that are involved in a traditional model of transactions may not be present. Therefore, while the underlying ideas of a variety of forms of governance are useful, the classic GVC governance taxonomies based on transaction cost economics are unsuitable.

This critique has already been alluded to within many studies of digital technologies and GVC, which have focused on modular components and standardised interfaces mediating between firms (as discussed above). Modular governance is the major governing mechanism used to incorporate the expansion of software and services within GVC. To elaborate, modular governance sits between pure markets and hierarchical relations, where production relations are defined by the high ability to codify complex requirements. These are embedded within standards, service interfaces and software. Within GVC, these types of configuration would reduce hierarchical relations in production, whilst allowing for complex requirements to be shared in interfirm production (for example, digital interfaces and industry platforms, amongst others, promote rich informational interfaces between transactors) (following Gereffi et al. 2005, Sturgeon 2002). For example, embedded within the recent notion of "massive modular" production, which covers multiple sectors and bundles of production, is a suggestion that these networks increasingly resemble modular governance (Sturgeon 2021, Thun et al. 2022).

This focus on modular governance is useful in that it cements a family of useful approaches to explore power, control and value that are relevant to AI and governance. One approach is to consider key "choke points" and/or centres of power within modular networks. Control of GVC then increasingly relates to firms (and places) which can control, monopolise or extract rent within the production chain (Foster 2024, Sturgeon 2021). More holistically, modular governance would also suggest a closer analysis of the emergence of modular standards and norms, and the way groups or individual firms seek to define standards and norms in GVC for power (Baglioni et al. 2021, Dallas et al. 2019).

However, it remains an open question if modular governance alone provides sufficient scope to understand the governance relationships linked to AI and openness. In the digital economy, differing conceptualisations of governance from other fields such as the platform studies, business ecosystems and open innovation have emerged. These approaches often centre on a more intense analysis of how technological resource owners, as a key governance actor, facilitate value creation in the digital economy by the way they configure and adapt resources and rules to shape interaction (Eaton et al. 2015, Van Alstyne et al. 2016). This somewhat resembles standards, but in also including private-oriented market systems, different forms of rules and regulations, "boundary resources" and affordances of technology, they capture some more dynamic aspects of governance relations (Eaton et al. 2015, Van Alstyne et al. 2016).

In sum, the literature on global value chains provides an important conceptualisation to think about the fragmentation of production networks and within these different notions of coordination, control and power. Here, three major areas have been elaborated. First, in terms of the underlying configuration of the value chain, relations mediate networks of software and services that impact the forms of the value chain. Bringing in openness further complicates these frameworks because it departs from traditional ideas of value relations within transactions to emphasise the need to consider a broader set of mediating factors, which will ultimately impact governance. Second, for leaders, there may be evolving dynamics of capitalism and value capture that move beyond the traditional forms discussed. Thirdly, and as a result of the above, we are liable to see heterogeneous types of governance patterns within downstream value chains. Drawing on this discussion of conceptual gaps, we will now turn to elaborating these ideas more concretely in the context of open software and AI.

## 4. The case of LLM and openness

We analyse the case of LLM (large language models) in this section. This form of AI, the most familiar and accessible models, which provide a chat interface to users, has been a major driver of AI popularisation. With LLM increasingly being multi-modal and with a major push towards "general purpose technologies" (GPT), they have attracted much interest and capital investment.

Analysis in this section is based on the rich technical literature of LLM. This provides detailed mapping of foundation LLM and the fragmentation of LLM production (Bommasani, Soylu, et al. 2023, Brown 2023, Hopkins et al. 2023, Widder & Nafus 2023). It also includes substantial discussion of different dimensions of openness in LLM (Liesenfeld et al. 2023, Solaiman 2023, Solaiman et al. 2019, Widder et al. 2024). With LLM becoming mainstream, literature also includes systematic audits of major LLMs, openness and AI supply chains in terms of different criteria of ethics and transparency (Bommasani, Klyman, et al. 2023, Ling et al. 2024, PAI 2021). As an example, the 2023 Foundations Model Transparency Index codifies 100 indicators in AI models linked to upstream (data, labour, compute), model (capabilities, model weights) and downstream (usage policies, distribution) properties of the AI (Bommasani, Klyman, et al. 2023). Although such work is more technically orientated, it is repurposed in this paper to provide a systematic understanding of LLM and openness, and to elaborate on the conceptual gaps outlined in the previous section.

## 4.1. Openness in LLM

| Model | Firm | Firm Model Type | Release Date | Licence |
|---|---|---|---|---|
| Bloom | Hugging Face (US) | Leading model | 2022/11 | OpenRAIL-M v1 ("Responsible AI" open licence) |
| LLaMa | Meta (US) | Leading model | v2: 2023/06 v3: 2024/04 | Custom: Free if you have under 700M users, cannot use LLaMA outputs to train other LLMs |
| Falcon | TII (UAE) | Leading model | 2023/05 | Apache 2.0 (Permissive open-source licence) |
| Mistral 7B | Mistral (EU) | Leading model | 2023/09 | Apache 2.0 (Permissive open-source licence) |
| Qwen1.5 | Alibaba (China) | Leading model | 2024/02 | Custom: Free if you have under 100M users and you cannot use Qwen outputs to train other LLMs |
| Gemma | Google (US) | Lightweight version of main model | 2024/02 | Free with restrictions. Models trained on Gemma outputs become subject to this license. |
| Grok-1 | xAI (formerly Twitter) (US) | Old version of the model | 2024/03 | Apache 2.0 |
| DeepSeek-V2 | DeepSeek (China) | Leading model | 2024/05 | Custom Free with usage restriction and models trained on DeepSeek outputs become DeepSeek derivatives, subject to this license. |
| Open AI (gpt-oss) | OpenAI (US) | Lightweight version of main model | 2025/08 | Apache 2.0 (Permissive open-source licence) |

*Table 1: Major open LLM. Data up to August 2025*
*Source: Adapted from (Yan 2025)*

Table 1 shows some of the dynamics around openness in major foundational LLM. During the earlier emergence of LLM, models were primarily released in proprietary forms (i.e. with users accessing models through online interfaces), but this has rapidly evolved. Most foundational AI firms still develop LLM in-house, but these are then opened to allow use within the wider community (Widder et al. 2024). Typically, in these scenarios, firms make *model weights* available (which are the learned numerical parameters that determine how the neural network processes model inputs). Subject to conditions as specified within the model licence, users can download and use these weights for their own application. Licences frequently permit external firms to commercially run models (often without a fee), host models within an infrastructure of their choice, integrate models with other services and "fine-tune" models (i.e. further adapting model weights through training on custom data). The most popular example of an open-weight LLM is Meta's LLaMA 3 model. It has a licence which allows modification of the model with retained ownership, redistribution with minimal acknowledgement and ownership of IP[8] (Meta 2025). Licencing Conditions on China's DeepSeek's LLM are more permissive, closely aligning with the "MIT licence" model, which has minimal restrictions. DeepSeek has also gone further by developing specific optimisations in model inference, which allows reduced compute capacity within a commercial implementation (DeepSeek-AI et al. 2025).

Although model weights are available, the literature highlights that there remain constraints around them being used downstream. Firstly, as shown in Table 1, there is variation in the licensing conditions of the model, which have real impacts on model users, including what they can do with the model and what types of commercial applications they can run. While some firms operate relatively flexible licences, other models include a range of restrictions, including maximum users for free commercial use, not permitting specific use cases (e.g. training new AI models using model outputs), or further conditions for commercialisation (e.g approval). As open source licences tend to operate as *ex post* enforcement (i.e. enforcement is mainly through action against non-implementation rather than ex ante agreements), licences may be relatively weak as a contracting mechanism. Nevertheless, licensing conditions mean control remains between the foundational firm and implementors.

Secondly, in the table we can also see that some LLM from major firms such as OpenAI, Google and xAI are more limited. Here, model weights are not available for the cutting-edge LLM models, but rather they are released with a delay, or of previous versions, or open model weights are a lightweight version of the leading-edge model (Bommasani, Klyman, et al. 2023, Solaiman et al. 2019). These less advanced models may potentially lead to implementors being less willing to use these models in a commercial context.

---

[8] Although there are additional licencing requirements for over 700 million monthly users and the licence does not permit use of the model for optimising competing ML models

Thirdly, and most importantly, while model weights are available, few firms provide full training data and details around creating the model[9]. This can be important if implementors want to stay at the leading edge, as they may not be able to recreate, adapt or understand LLM[10]. Without access to the details and in an environment of closed data and high costs of compute, AI implementors may end up in dependency on foundational firms, even as the openness of LLM is expanding (Liesenfeld et al. 2023). It is these types of dependency that have been at the heart of the "open washing" argument[11],

> "Instead of providing meaningful documentation and access, they [open models] are effectively wrappers around closed models, inheriting undocumented data, failing to provide annotated reinforcement learning from human feedback (RLHF) training data and labour-process information and rarely publishing their findings, let alone documenting these in independently reviewed publications" *(Widder et al. 2024 p.829)*.

Bringing these three different aspects of constraints together. Based on auditing of LLM, Solamiman(2023) suggests we can think of a gradient framework for "model release" (openness), as outlined in **Error! Reference source not found.**. The framework moves from fully proprietary LLM on the left, through LLM which are released under a gradual or staged release, to "gated models" with some more flexible access (e.g. via an API), to fully open models in terms of model weights and data on the right. As shown in his 2023 implementation, this framework is useful to reinforce that there is a diversity of strategies being deployed around openness and AI by different firms.

---

[9] There are exceptions to this, but this tends to be within smaller AI models led by universities, governments or those with more social goals (such as the BLOOM multiligual model). In these cases, the focus may be on providing more transparency sharing of models (Bommasani, Klyman, et al. 2023, Khanal et al. 2024)

[10] Models often provide some information about how their models have been built. Firms often release a technical paper for a new model, outlining their innovations and benchmark results on release, more recently model releases is often accompanied by a "model card". This can include a more or less detailed outlines of data sources and the types of cleaning and training undertaken but is typically insufficient to replicate these models

[11] Although given some the broader ethical discussion around flagship AI "getting into the wrong hands" some of the leading firms have positioned stricter licencing as being about acting as a responsible gatekeepubg to appropriate applications of their models (Solaiman 2023)

*Figure 2: Gradient of AI access. Models listed are based on analysis as of 2023*
*Source: (Solaiman 2023 p.114)*

## 4.2. Strategic market openness

The previous section identified variation in terms of the openness of LLM. We can then think about this variation in terms of capitalist dynamics of foundational firms, through tracking the discussion of firms around the goal of openness. For major firms that only release non-cutting edge or lightweight model weights, it may be that these can only be used in limited application cases. They may be used within research or product development, but implementors defer back to using major platforms or full licences in implementation[12]. Overall, the dynamics here, in line with the longer-running philosophies around "Free" within digital technology, are that old or lightweight models are cross-subsidies or "freemium" versions that attract users to platforms (see Anderson 2010 for a classic outline).

For firms that push more open models, there can still be an advantage driven by market development and competitiveness. By making models and services openly available and ensuring adoption at an earlier stage, firms may look to support competition in LLM. For example, open LLM weights have been seen as one important factor driving the reduction of costs for LLM through competition (McKinsey 2025). A closely related justification is that opening AI can be seen as a strategic direction to ensure a more shared dynamic of AI processes and resources (rather than it being defined in proprietary standards or firm dominance). It is these types of articulation that have accompanied Meta's business model centres on openness as discussed by the CEO Mark Zuckerberg,

---

[12] To elaborate, even if lightweight LLM themselves can serve some application scenarios, frequently the associated platform/service offerings of leading firms will include leading-edge optimisations and large-scale compute improvements which smaller implementors will unable to replicate on open LLM. As such in the current moment, the attraction using more property model remains very strong.

> *"….open-source software often becomes an industry standard, and when companies standardize on building with our stack, that then becomes easier to integrate new innovations into our products" (Gent 2024)*

In contrast to a firm like OpenAI, where having the leading-edge AI model is the *business goal*, this quote highlights how Meta may also see advanced AI as a *tool* for advancing other fundamental business goals (i.e. intermediating between connections and building rich social networks).

Beyond the dynamics of standardisation, having the ability to guide standards can be important for foundational firms in terms of interfacing with firm infrastructure. Widder et al(2024) illustrate this within the case of the earlier *TensorFlow* from Google (Alphabet). Here, openness was beneficial because it drove customer use of Google's dedicated AI server infrastructure,

> "TensorFlow has been created to easily and intuitively operate with Google's Tensor Processing Unit (TPU) hardware, the powerful proprietary computing infrastructure core to Google's cloud AI computing business" *(Widder et al. 2024 p.831)*

For some more critical analysts, such approaches may lead to software lock-ins. As Widder continues, it "…*leads developers to create AI systems that, like Lego, snap into place with…company systems"* (Widder et al. 2024 p.830).

Beyond capitalist motivations for open models, dynamics may also move towards broader goals. In Table 1, this is shown with the earlier BLOOM model, which followed an academic "Big Science" type approach of crowdsourcing to build the model. Similarly, the Falcon model, which excels in Arabic texts, emerged from the UAE's Technology Innovation Institute, an academic institute supported by the Kingdom's public research bodies in partnership with academia. In China, the rapidly expanding popularity of open models such as DeepSeek and Qwen LLM may align with Chinese firm strategies for competitiveness in the context of trade restraints.

In sum, the motivations for opening AI components are rather different to previous capitalist dynamics. Combining these explanations into a broader umbrella, *strategic market openness* describes a family of approaches where openness in high-value components closely aligns with capitalist or non-capitalist goals of AI firms.

## 5.     Openness and implications in downstream value chains

The above discussion provides a basis to consider the emergence of AI openness as a diversity of approaches, and of *strategic market openness* being a key driver. Based on this analysis, this section reflects how these varying notions then link into governance. Developing the original schematic of a downstream value chain from Figure 1, a variety of dynamics of foundation firms are incorporated into a framework, with the resultant governance discussed below.

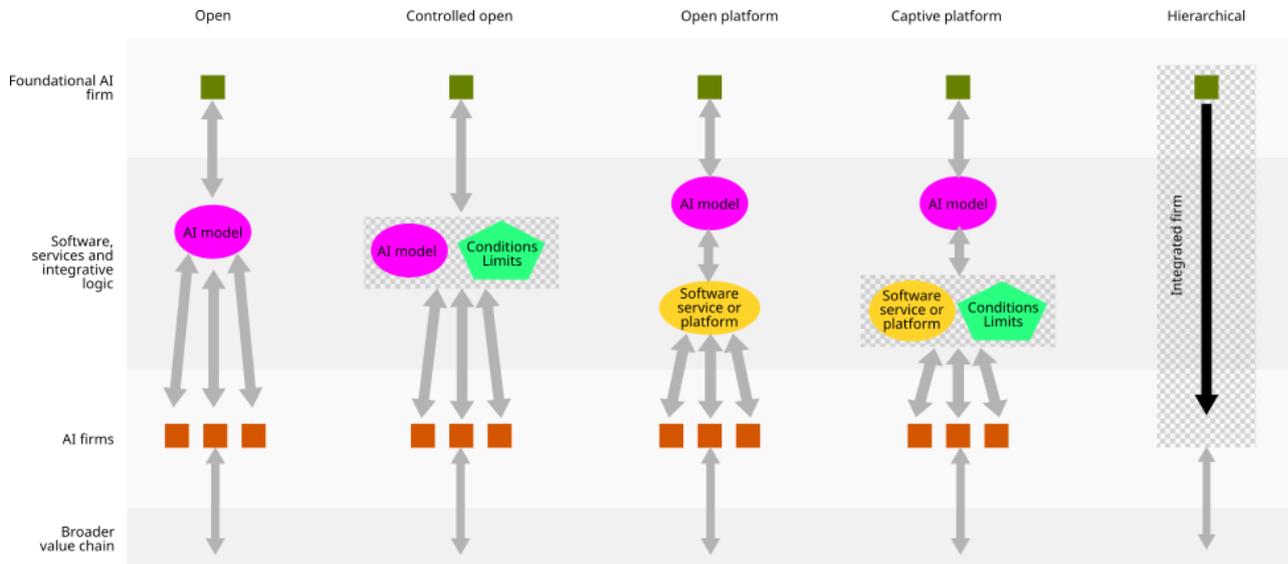

*Figure 3: Governance models in AI value chains. Source: Author's elaboration.*

In the above framework, the openness of foundational firms will impact the fragmentation of downstream value chains. Theories from modular production networks, platform governance and open innovation support thinking about the new ways that foundational firms can govern downstream networks:

**1) Hierarchical:** On the right of Figure 3 are "hierarchical" models of governance that most resemble a more conventional (closed source) software services or off-the-shelf software driven by contracts. Large in specific areas may use their own internal data, capabilities and existing compute to develop foundation models or integrate them within more conventional software/service value chains. In this scenario, the majority of major innovations would be internalised within the firm. For example, well-established firms such as Siemens and IBM are moving into AI and have developed applications around predictive analytics, real-time monitoring and optimisation. Although resultant software and services may provide the ability for firms to customise and develop solutions[13], often the nature of software use is top-down. For governance, this would imply that as AI becomes more central to value, in some sectors implementors are likely to be pulled into a captive relationship with the suppliers, shaping the nature, risk and competitiveness in a sector.

---

[13] For example, IBM's Watsonx provides a platform for industrial device development which may better align to the platform type as per the next sub-section.

**2) Captive externalised governance:** This aligns closely to platform governance or modularised service ecosystem paradigm as applied to AI (Tiwana 2014, Van Alstyne et al. 2016). AI models may make claims to openness due to being extendable or modularizable. For example, Google's AI platforms offer a range of services, ranging from its Gemini LLM, natural-language processing models, AI data processing, as well as third-party services. Typically, AI implementors would access such functionality through software APIs, paying a per-process or subscription fee for services. The setup of AI access is, therefore, quite flexible and allows implementors to gain access to a range of AI functionalities within well-documented services. But there are likely to under certain constraints on how an implementor is able to use this service, including around intellectual property, usage, costs, types of application, amongst others. This mode of AI may also extensively use cross-subsidisation, such as by making the weights of lightweight or older models available to download under open licences (as discussed previously). However, given the conditions and behind-the-leading-edge, AI implementors are likely to prefer to remain under a platform-like model.

In captive externalised scenarios, many of the complications of AI will remain opaque to the user. For example, optimisation of AI computation and even the specific details of what models are in use may not be visible to the implementor. This may be beneficial in terms of implementors having the agility to adopt AI rapidly, supported by the substantial "scaffold" of services, utilities and optimisations. The term "captive externalised" is used in terms of governance because this opacity results in foundation firms maintaining control of key aspects of compute, models and data, with users still pulled into relatively captive relations, limited by the service offering and specificities of the (rapidly changing) platform. Mirroring platform studies, captive externalised governance also poses risks of "platform lock-in", increasing service fees and industry concentration.

**3) Open Platform:** Given that effective AI foundational firms need to tie together a range of resources (data, computing, software, cloud) (Mueller 2025), some leading firms in the AI space have centred their focus on "open platforms", particularly related to the provision of the infrastructural aspect of AI. An example of this would be an infrastructural firm such as Amazon. Although Amazon only offers more limited AI models, it offers a range of more open models (from competing firms) within its own infrastructure (e.g. Deepseek and Meta models). This works alongside the ability to closely integrate with existing data and service resources efficiently. For example, Amazon allows single-click integration with other AWS applications, libraries and services (van der Vlist et al. 2024). Without being wedded to a single model or service provider, open platforms provide more customisability and ability for implementors to rapidly switch across AI providers. In such cases, however, there remain some significant aspects of control. Specifically, infrastructure providers tend to control key "choke points" of AI production that cement their control. This mainly revolves around optimised infrastructure access in the current moment, which limits the ability for implementers to run such models elsewhere[14]. This still leads to quite captive relationships because computing power to build and train AI models is a major constraint for implementing AI (or requires more capabilities and capital). With industry concentration around AI data centres, this might still constrain firms from moving outside these relationships.

**4) Controlled openness:** Leading foundational AI firms have released models close to leading-edge under relatively promissory licences such as foundation firms like Meta and DeepSeek. These provide far fewer constraints to users of AI, with potential for commercial use, hosting modification and exploration. Nevertheless, use is subject to certain policy that limits some uses and applications (albeit not contracted in the traditional sense), as well as more inherent limits due to open weights, such as lack of data or compute, that can prevent AI implementors from creating their own version of these models. From the perspective of these foundation firms, more flexible governance allows a modicum of control over the model, but the main goal is to support adoption to be able to more strongly define plurality and/or shape standardisation processes and norms within the nascent AI field.

**5) Openness:** On the left of the framework are cases where foundational AI firms move most closely towards formal definitions of open source AI: where full models, tool chains and infrastructure might be put into the public domain under highly flexible licensing conditions. There are limited examples of lead AI firms following this, but some lead firms, such as DeepSeek, are moving in this direction. There are also initiatives within university-driven AI models, such as BLOOM, which although more dated, align with these goals. Here, we might more strongly ascribe a very different perspective, such as overcoming geopolitics, ensuring sovereignty and broader academic goals around open knowledge.

This framework, with the different types of governance, is summarised in Table 2. It represents a useful way to define different types of governance, but it will require further empirical analysis, as AI is used at scale across different sectors. There remain questions about how these more indirect types of power are manifested downstream across different stages of value chains.

---

[14] Infrastructure lock-in could also come from how firms use customers, model weights or proprietary data as choke points in the future.

Therefore, in line with GVC methodologies, as openness in AI expands, particularly for LLM, there is potential to flesh out this taxonomy through more detailed case studies.

| Governance | Explanation | Capitalist dynamics | Control vs Capability trade-off | AI examples |
|---|---|---|---|---|
| **Hierarchical** | Foundational firms providing specific services or software to users that incorporate AI. | **Internalising competencies**<br>• AI is seen as a core competency in the organisation. | More hierarchical control – most AI competencies remain in-house, and AI users will use it under a service contract. | AI use in major industries (health, transport, industry, etc.)<br>• Procore (construction)<br>• Tractable (insurance software)<br>• John Deere (agribusiness) |
| **Captive platform** | Firms provide access to AI resources through clouds, API or platforms. | **Platform governance**<br>• Firms keep key resources around AI internal to ensure value capture (e.g. data within platforms) | Flexible use for AI users on platforms and with beneficial cross-subsidies<br><br>Platforms control key scenarios and resources with potential platform "lock-in" | Leading foundational models.<br>• OpenAI GPT 5, 4, 3.5<br>• Google Gemini models |
| **Open Platform** | Firms mainly provide AI infrastructure and facilitate AI functionality. | **Open platform control**<br>• Firms maintain control of "chokepoints" as a source of value (e.g. infrastructure) | Higher skill requirements for AI over captive platform, but flexible use, including competitor models and libraries<br><br>Key AI constraints, such as data, efficient compute, lead to control | Leading AI infrastructure firms<br>• Amazon Bedrock<br>• Hugging Face<br>• Other software platforms which incorporate AI plugins (ERP, cloud, SAAS) |
| **Controlled openness** | Foundational firms release AI under some conditions and constraints | **Standardisation**<br>• Lead firms see AI as a *tool* to capture value and seek to promote standards and a broad market, | Very high skill requirements to use (e.g. setting up servers, fine-tuning) but larger flexibilities.<br><br>AI users remain in a dependent relation (e.g model parameters, licence conditions) | Foundational firms with model weight available<br>• Deepseek<br>• LLaMA model<br>• Mistral model<br>• QWEN model |
| **Openness** | Firms create open-source AI resources and provide them under permissive licences. | **Beyond solely economic dynamics**<br>• Goals may include academic goals, sovereignty. geopolitical<br>• Open-source AI systems as an end in itself for developers | Very high skill requirements to use (e.g. setting up servers, fine-tuning) but larger flexibilities. | Limited cases:<br>• BLOOM (big science initiative)<br>• KISH (EU AI model – in development)<br>• EleutherAI models (specific open source dedicated model |

*Table 2: Summary of governance types and openness in AI value chains. Source: Author's elaboration.*

# 6.     Conclusion

The move towards openness in AI has been a significant trend, prompting debate in the field. Extending work that has incorporated chain approaches into the analysis of AI, this paper used openness as an explanatory mechanism to analyse the ongoing fragmentation of AI. Specifically, the work then extends the analysis of AI chains through a closer and critical application of theories of global value chains. Empirical analysis supports this discussion based on technical analysis and auditing of openness surrounding "fundamental" LLM models of leading firms.

The GVC literature on service, software and the digital economy allows us to highlight the heterogeneity of firms in terms of different mixes and approaches to the provision of openness. This work then aligns with the literature that suggests broadening "governance", incorporating notions from modular production networks, platform governance and open innovation that allow us to sketch out the implications of openness in terms of control and capabilities for firms downstream in AI value chains.

Adapting GVC/GPN discussion around concepts of capitalist dynamics of lead firm, we outline the notion "strategic market openness" to explain scenarios where openness enhances major value capture strategies of lead firms (including data, infrastructure, standards control and geopolitics). A key conclusion here would be that trends of strategic market openness are not just a fleeting moment. Firms building foundational models have sustainable motivations to follow this direction, including, in some cases, a move towards more "open source" implementations of AI, even if this implies making huge capital investments available for free and for modification for downstream implementors.

Through the framework of openness and governance, the paper emphasised heterogeneous types of governance. Hence, it argues that there is likely to be a variation in the balance between coordination, control, power vs flexibility, capability building and upgrading potential according to different types of foundational AI firm strategy (see right column in Table 2, where some elaboration of this is made for each governance type)

Foundational AI firms with their unprecedented capital investments in AI are unlikely to relinquish their lead positions.  In downstream AI value chains, even when open governance occurs, platformization, cross-subsides, control of "choke points", infrastructure and standardisation all represent contemporary dimensions of control. Claims of openness by foundational AI firms, therefore, even in more open scenarios, are not contradictory to potentially quite powerful lead AI firms.

Notwithstanding these debates on governance and power, we suggest that openness in AI can be a significant step in terms of supporting expanded use and capabilities in AI. Whilst a fuller empirical analysis of openness and its link to governance will need to be developed further, the idea that complete toolchains for AI, including leading AI LLM models, huge datasets, infrastructure and libraries used by foundational firms to build and implement leading models, are all available to non-lead firms should not be neglected. What is being made available here is innovative software and services, often subject to capital-intensive investments and often freely available.

While the expansion of AI is still playing out, this work should be part of the consideration for future scenarios. Conceivably, we will see computation costs for training and implementation costs; and where fine-tuning of AI can be done on less and/or synthetic data; and where a wider range of applications for AI emerge. In such contexts, open models and tools are likely to become more available and viable for use. Therefore, this dynamic has important implications for capability building, specialisation and/or upgrading in GVC. The wealth of AI suggests resources which can rapidly be downloaded, adapted, including local development, with clear resources and documentation from leading actors. At least when considering these shifts in terms of the classic GVC and industrial development literature on capabilities, it would suggest future new opportunities (Lall 1992, Pietrobelli & Rabellotti 2011, Whitfield & Staritz 2021).

# References

Anderson, C. (2010) *Free: How Today's Smartest Businesses Profit by Giving Something for Nothing*. Random House Business, London.

Anwar, M.A. (2025) Value Chains of AI, in *The Future of Labour: How AI, Technological Disruption and Practice Will Change the Way We Work*, A. Larsson & A. Hatzigeorgiou (eds), Routledge, London.

Attard-Frost, B. & Widder, D.G. (2025) The Ethics of AI Value Chains. *Big Data & Society*, 12(2).

Baglioni, E., Campling, L. & Hanlon, G. (2021) Beyond Rentiership: Standardisation, Intangibles and Value Capture in Global Production. *Environment and Planning A: Economy and Space*, 55(6), pp. 1528–1547.

Baines, T.S., Lightfoot, H.W., Benedettini, O., et al. (2009) The Servitization of Manufacturing: A Review of Literature and Reflection on Future Challenges. *Journal of Manufacturing Technology Management*, 20(5), pp. 547–567.

Bair, J. (2005) Global Capitalism and Commodity Chains: Looking Back, Going Forward. *Competition and Change*, 9(2), pp. 153–180.

Benkler, Y. (2006) *The Wealth of Networks: How Social Production Transforms Markets and Freedom*. Yale University Press, London.

Bommasani, R., Hudson, D.A., Adeli, E., et al. (2022) On the Opportunities and Risks of Foundation Models. Available from: http://arxiv.org/abs/2108.07258 [Accessed 5 August 2025].

Bommasani, R., Klyman, K., Longpre, S., et al. (2023) The Foundation Model Transparency Index. Available from: http://arxiv.org/abs/2310.12941 [Accessed 13 December 2024].

Bommasani, R., Soylu, D., Liao, T.I., et al. (2023) Ecosystem Graphs: The Social Footprint of Foundation Models. Available from: https://arxiv.org/abs/2303.15772 [Accessed 19 December 2024].

Bratton, B.H. (2016) *The Stack: On Software and Sovereignty*. 1 edition. The MIT Press, Cambridge, Massachusetts.

Brown, I. (2023) *Allocating Accountability in AI Supply Chains*, Ada Lovelace Institute, London, UK.

Chesbrough, H.W. (2006) *Open Innovation: The New Imperative for Creating and Profiting from Technology*. Harvard Business Press, Harvard, MA.

Cobbe, J., Veale, M. & Singh, J. (2023) *Understanding Accountability in Algorithmic Supply Chains*. Paper presented at: New York, NY, USA, 12th Jun. Available from: https://dl.acm.org/doi/10.1145/3593013.3594073 [Accessed 20 December 2024].

Crawford, K. (2021) *Atlas of AI: Power, Politics, and the Planetary Costs of Artificial Intelligence*. Yale University Press.

Dallas, M.P., Ponte, S. & Sturgeon, T.J. (2019) Power in Global Value Chains. *Review of International Political Economy*, 26(4), pp. 666–694.

DeepSeek-AI, Guo, D., Yang, D., et al. (2025) DeepSeek-R1: Incentivizing Reasoning Capability in LLMs via Reinforcement Learning. Available from: http://arxiv.org/abs/2501.12948 [Accessed 2 February 2025].

Dicken, P. (2011) *Global Shift: Transforming the World Economy*. 6th ed. Guildford Press, London, UK.

Dupuis, M. (2024) Algorithmic Management and Control at Work in a Manufacturing Sector: Workplace Regime, Union Power and Shopfloor Conflict over Digitalisation. *New Technology, Work and Employment*, 40(1), pp. 81–101.

Eaton, B., Elaluf-Calderwood, S., Sorensen, C., et al. (2015) Distributed Tuning of Boundary Resources: The Case of Apple's iOS Service System. *MIS Quarterly: Management Information Systems*, 39(1), pp. 217–243.

Ferrari, F. (2023) Neural Production Networks: AI's Infrastructural Geographies. *Environment and Planning F*, 2(4), pp. 459–476.

Foster, C. (2024) Theorizing Globalized Production and Digitalization: Towards a Re-Centering of Value. *Competition & Change*, 28(1), pp. 189–208.

Foster, C. & Graham, M. (2017) Reconsidering the Role of the Digital in Global Production Networks. *Global Networks*, 17(1), pp. 66–88.

Foster, C., Graham, M., Mann, L., et al. (2018) Digital Control in Value Chains: Challenges of Connectivity for East African Firms. *Economic Geography*, 94(1), pp. 68–86.

Foster, C.G. & Azmeh, S. (2023) *Aligning Digital and Industrial Policy to Foster Future Industrialization*, Policy Brief: Insights on Industrial Development, 4, UNIDO, Vienna, Austria.

Gambacorta, L. & Shreet, V. (2025) *The AI Supply Chain*, 154, Bank for International Settlements, Basel, Switzerland.

Gent, E. (2024) *The Tech Industry Can't Agree on What Open-Source AI Means. That's a Problem.*, MIT Technology Review.

Gereffi, G. (2001) Beyond the Producer-Driven/Buyer-Driven Dichotomy The Evolution of Global Value Chains in the Internet Era. *IDS Bulletin*, 32(3), pp. 30–40.

Gereffi, G. & Fernandez-Stark, K. (2010) *The Offshore Services Value Chain: Developing Countries and the Crisis*. The World Bank.

Gereffi, G., Humphrey, J. & Sturgeon, T. (2005) The Governance of Global Value Chains. *Review of International Political Economy*, 12(1), pp. 78–104.

Gurumurthy, A., Bharthur, D., Chami, N., et al. (2020) *Unskewing the Data Value Chain– A Policy Research Project for Equitable Platform Economies: Background Paper*, IT for Change.

Havice, E. & Pickles, J. (2019) On Value in Value Chains, in *Handbook on Global Value Chains*, S. Ponte, G. Gereffi, & G. Raj-Reichert (eds), Edward Elgar Publishing, Cheltenham, UK, pp. 169–182.

Heeks, R. & Spiesberger, P. (2024) *Constructing an AI Value Chain and Ecosystem Model*, Digital Development Working Paper Series, 109, Centre for Digital Development, University of Manchester, Manchester, UK.

Henderson, J., Dicken, P., Hess, M., et al. (2002) Global Production Networks and the Analysis of Economic Development. *Review of International Political Economy*, 9(3), pp. 436–464.

Herpig, S. (2020) *Understanding the Security Implications of the Machine-Learning Supply Chain*, Interface, Berlin, Germany.

Hippel, E. von & Krogh, G. von (2003) Open Source Software and the 'Private-Collective' Innovation Model: Issues for Organization Science. *Organization Science*, 14(2), pp. 209–223.

Hopkins, A., Struckman, I., Madry, A., et al. (2024) AI Supply Chains Aren't AI Value Chains. *Thoughts on AI Policy*. Available from: https://aipolicy.substack.com/p/supply-chains-6 [Accessed 20 December 2024].

Hopkins, A., Struckman, I., Madry, A., et al. (2023) The Diverse Landscape of AI Supply Chains: The AIaaS Supply Chain Dataset. *Thoughts on AI Policy*. Available from: https://aipolicy.substack.com/p/supply-chains-3-5 [Accessed 20 December 2024].

Horner, R. & Nadvi, K. (2018) Global Value Chains and the Rise of the Global South: Unpacking Twenty-First Century Polycentric Trade. *Global Networks*, 18(2), pp. 207–237.

Humphrey, J. (2018) *Value Chain Governance in the Age of Platforms*, 714, IDE-JETRO, Chiba, Japan.

Hung, K.-H. (2024) Artificial Intelligence as Planetary Assemblages of Coloniality: The New Power Architecture Driving a Tiered Global Data Economy. *Big Data & Society*, 11(4).

IHCAI (2023) *Governing Open Foundation Models*, Issue Brief HAI Policy & Society, Institute on Human-Centered Artificial Intelligence, Stanford University, Stanford, MA.

Kaplinsky, R. & Morris, M. (2001) *A Handbook for Value Chain Research*, IDRC, Ottawa, Canada.

Khanal, S., Zhang, H. & Taeihagh, A. (2024) Building an AI Ecosystem in a Small Nation: Lessons from Singapore's Journey to the Forefront of AI. *Humanities and Social Sciences Communications*, 11(1), pp. 1–12.

Krzywdzinski, M. (2017) Automation, Skill Requirements and Labour-Use Strategies: High-Wage and Low-Wage Approaches to High-Tech Manufacturing in the Automotive Industry. *New Technology, Work and Employment*, 32(3), pp. 247–267.

Küspert, S., Moës, N., & Connor Dunlop (2023) *The Value Chain of General-Purpose AI*, Ada Lovelace Institute, London, UK.

Lall, S. (1992) Technological Capabilities and Industrialization. *World Development*, 20(2), pp. 165–186.

Larsson, S. & Heintz, F. (2020) Transparency in Artificial Intelligence. *Internet Policy Review*, 9(2), pp. 1–16.

Lee, K.-F. (2018) *AI Superpowers China, Silicon Valley, and the New World Order*. Houghton Mifflin Harcourt, Boston.

Lessig, L. (2004) *Free Culture: How Big Media Uses Technology and the Law to Lock Down Culture and Control Creativity*. Penguin, London, UK.

Liesenfeld, A., Lopez, A., View Profile, et al. (2023) Opening up ChatGPT: Tracking Openness, Transparency, and Accountability in Instruction-Tuned Text Generators. *Proceedings of the 5th International Conference on Conversational User Interfaces*, pp. 1–6.

Ling, L., Liling, H., Wenyang, F., et al. (2024) *Transparency Assessment of 15 Chinese Large Models: Only 4 Allow Users to Withdraw Voiceprint Data*

*(15款国产大模型透明度测评：仅4款允许用户撤回声纹数据)*, Nandu Digital Economy

Governance Research Center [南都数字经济治理研究中心], Nandu , China.

Low, P. (2013) *The Role of Services in Global Value Chains*, The Fung Global Institute Working Paper Series, FGI-2013-1, Fung Global Institute, Hong Kong.

McKinsey (2025) *How Open Source AI Solutions Are Reshaping Business*, McKinsey.

Meta (2025) Meta Llama 3 License [Online]. Available from: https://www.llama.com/llama3/license/ [Accessed 10 April 2025].

Mills, S. (2024) Algorithms, Bytes, and Chips: The Emerging Political Economy of Foundation Models. Available from: https://papers.ssrn.com/abstract=4834417 [Accessed 6 July 2024].

Mueller, M.L. (2025) It's Just Distributed Computing: Rethinking AI Governance. *Telecommunications Policy*, 49(3).

OSI (n.d.) The Open Source AI Definition – 1.0 [Online], *Open Source Initiative*. Available from: https://opensource.org/ai/open-source-ai-definition [Accessed 9 April 2025].

PAI (2021) *Responsible Sourcing of Data Enrichment Services*, Partnership on AI, San Francisco, CA.

Pietrobelli, C. & Rabellotti, R. (2011) Global Value Chains Meet Innovation Systems: Are There Learning Opportunities for Developing Countries? *World Development*, 39(7), pp. 1261–1269.

Ponte, S. & Gibbon, P. (2005) Quality Standards, Conventions and the Governance of Global Value Chains. *Economy and Society*, 34(1), pp. 1–31.

Ponte, S. & Sturgeon, T. (2014) Explaining Governance in Global Value Chains: A Modular Theory-Building Effort. *Review of International Political Economy*, 2(1), pp. 1–29.

Prahalad, C.K. & Hamel, G. (1990) The Core Competence of the Corporation. *Harvard Business Review*, (May–June), pp. 3–22.

Solaiman, I. (2023) *The Gradient of Generative AI Release: Methods and Considerations*. Paper presented at: New York, NY, USA, 12th Jun. Available from: https://dl.acm.org/doi/10.1145/3593013.3593981 [Accessed 13 December 2024].

Solaiman, I., Brundage, M., Clark, J., et al. (2019) Release Strategies and the Social Impacts of Language Models. Available from: http://arxiv.org/abs/1908.09203 [Accessed 20 December 2024].

Srnicek, N. (2020) *Data, Compute, Labour*, Ada Lovelace Institute, London, UK.

Stalder, F. (2018) *The Digital Condition*. Polity, Cambridge, UK ; Medford, MA.

Sturgeon, T.J. (2021) Upgrading Strategies for the Digital Economy. *Global Strategy Journal*, 11(1), pp. 34–57.

Sturgeon, T.J. (2002) Modular Production Networks: A New American Model of Industrial Organization. *Industrial and Corporate Change*, 11(3), pp. 451–496.

Thun, E., Taglioni, D., Sturgeon, T., et al. (2022) *Massive Modularity: Understanding Industry Organization in the Digital Age: The Case of Mobile Phone Handsets*, World Bank, Policy Research Working Paper, 10164, The World Bank.

Tiwana, A. (2014) *Platform Ecosystems: Aligning Architecture, Governance, and Strategy*. Newnes, Waltham, MA.

Valdivia, A. (2024) The Supply Chain Capitalism of AI: A Call to (Re)Think Algorithmic Harms and Resistance through Environmental Lens. *Information, Communication & Society*, 0(0), pp. 1–17.

Van Alstyne, M.W., Parker, G.G. & Choudary, S.P. (2016) Pipelines, Platforms, and the New Rules of Strategy. *Harvard Business Review*, 94(4), pp. 54–62.

van der Vlist, F., Helmond, A. & Ferrari, F. (2024) Big AI: Cloud Infrastructure Dependence and the Industrialisation of Artificial Intelligence. *Big Data & Society*, 11(1), p. 20539517241232630.

Von Hippel, E. (2001) Innovation by User Communities: Learning from Open-Source Software. *MIT Sloan Management Review*, 42(4), p. 82.

Whitfield, L. & Staritz, C. (2021) The Learning Trap in Late Industrialisation: Local Firms and Capability Building in Ethiopia's Apparel Export Industry. *The Journal of Development Studies*, 57(6), pp. 980–1000.

Widder, D.G. & Nafus, D. (2023) Dislocated Accountabilities in the "AI Supply Chain": Modularity and Developers' Notions of Responsibility. *Big Data & Society*, 10(1), pp. 1–12.

Widder, D.G., Whittaker, M. & West, S.M. (2024) Why 'Open' AI Systems Are Actually Closed, and Why This Matters. *Nature*, 635(8040), pp. 827–833.

Woodcock, J. (2020) The Algorithmic Panopticon at Deliveroo: Measurement, Precarity, and the Illusion of Control. *Ephemera: Theory & Politics in Organizations*, 20(3), pp. 67–95.

Wooldridge, M. (2021) *A Brief History of Artificial Intelligence: What It Is, Where We Are, and Where We Are Going*. Flatiron Books, New York.

Yan, E. (2025) A List of Open LLMs Available for Commercial Use. [Online], *GitHub*. Available from: https://github.com/eugeneyan/open-llms?tab=readme-ov-file [Accessed 11 August 2025].

Yeung, H.W.C. & Coe, N. (2015) Towards a Dynamic Theory of Global Production Networks. *Economic Geography*, 91(1), pp. 29–58.

Yoo, Y., Henfridsson, O. & Lyytinen, K. (2010) Research Commentary—The New Organizing Logic of Digital Innovation: An Agenda for Information Systems Research. *Information Systems Research*, 21(4), pp. 724–735.

Zhang, M.M., Cooke, F.L., Ahlstrom, D., et al. (2025) The Rise of Algorithmic Management and Implications for Work and Organisations. *New Technology, Work and Employment*, Early view.

Zysman, J., Murray, J., Feldman, S., et al. (2011) *Services with Everything: The ICT-Enabled Digital Transformation of Services*, BRIE Working Paper, 187a, UC Berkeley, Berkeley, CA.